\documentclass[twocolumn,showpacs,preprintnumbers,prb,amsmath,amssymb]{revtex4}
\usepackage{graphicx}% Include figure files
\usepackage{dcolumn}% Align table columns on decimal point
\usepackage{bm}% bold math

\begin{document}

%\preprint{DV/LiCo}

\title{Magnetic field induced orientation of superconducting MgB$_2$ crystallites determined by X-ray diffraction}% Force line breaks with \\

\author{J. Li,$^1$ D. Vaknin\footnote{Electronic address: vaknin@ameslab.gov},$^1$ S. L. Bud'ko,$^1$ P. C. Canfield,$^1$
D. Pal,$^2$  M. R. Eskildsen,$^2$ Z. Islam,$^3$ V. G. Kogan$^1$}
 %\altaffiliation[Also at ]{}%Lines break automatically or can be forced with \\
%\author{}%
%\email{vaknin@ameslab.gov}
\affiliation{$^1$Ames Laboratory and Department of Physics and Astronomy, Iowa State University, Ames, Iowa 50011, USA\\
$^2$ Department of Physics, University of Notre Dame, Notre Dame, Indiana 46556, USA\\
$^3$ Advanced Photon Source, Argonne National Laboratory, Argonne, Illinois 60439, USA
}%

%\affiliation{
%This line break forced% with \\}%

\date{\today}% It is always \today, today,
\begin{abstract}
X-ray diffraction studies of fine polycrystalline samples of MgB$_2$ in the superconducting state reveal that crystals orient with their \emph{c}-axis in a plane normal to the direction of the applied magnetic field.  The MgB$_2$ samples were thoroughly ground to obtain average grain size 5 - 10 $\mu$m in order to increase the population of free single crystal grains in the powder.  By monitoring Bragg reflections in a plane normal to an applied  magnetic field we find that the powder is textured with significantly stronger (\emph{0,0,l}) reflections in comparison to (\emph{h,k,0}), which remain essentially unchanged.    The orientation of the crystals with the \emph{ab}-plane  parallel to the magnetic field at all temperatures below $T_c$ demonstrates that the sign of the torque under magnetic field does not alter, in disagreement with current theoretical predictions.
%\verb+\pacs{#1}+ command.
\end{abstract}

\maketitle %%
\section{Introduction}
 It is well established experimentally and theoretically that the layered two-band superconductor MgB$_2$ exhibits unique and intriguing anisotropic properties.  These are manifested in measurements of the ratios of the upper critical fields $\gamma_H = H^{\perp c}_{c2}/H^{\parallel c}_{c2}$ for magnetic fields along the $c-$axis and the $ab-$plane and of the London penetration depth $\gamma_{\lambda} = \lambda_c/\lambda_{ab}$.\cite{Papavassiliou2001, Simon2001, Bud'ko2001b, Golubov2002, Zehetmayer2002, Bud'ko2002, Angst2002, Kogan2002, Cubitt2003, Kogan2003, Rydh2004, Angst2004, Lyard2004, Feltcher2005}  The preponderance of experiments are in agreement with regard to the temperature dependence of $\gamma_H$, which shows increase  from $\gamma_H \approx 2.0$ around $T_c$ to $\gamma_H \approx 6 $ at low temperatures.\cite{Angst2002, Bud'ko2002, Feltcher2005}  However, the temperature dependence of $\gamma_{\lambda}$, in particular under applied magnetic field, is not settled yet.\cite{Lyard2004}  Experimental evidence shows that $\gamma_\lambda = \gamma_H \approx 2$ around $T_c = 40$ K, and as the temperature decreases it approaches the isotropic value $\gamma_\lambda = 1$ (see Fig. 4 in Ref. \cite{Feltcher2005}).  Some experimental studies claim that the $\gamma_H(T,H)=\gamma_\lambda(T,H)$.\cite{Angst2004,Lyard2004}  For a crystal with $\gamma_{\lambda} = \gamma_H$ in a field along the $c$-axis, the vortex energy is given by $\sim(\phi_0/4\pi\lambda_{ab})^2$ within a factor of order unity, where $\phi_0$ is the flux quantum.  The energy is approximately $\gamma_{\lambda}$ times lower when the field is perpendicular to the $c-$axis: $\sim(\phi_0/4\pi)^2/(\lambda_{ab}\lambda_c) =  (\phi_0/4\pi\lambda_{ab})^2/\gamma_{\lambda}$.  As a result, superconducting free single crystal grains placed in the magnetic field experience a torque that tends to orient their \emph{ab} plane parallel to the field.  Such behavior has been observed in YBCO by neutron scattering experiments\cite{Tranquada1988} on powders, and in single crystal torque measurements\cite{Farrell1988} performed at temperatures close to $T_c$.

 Recent theoretical predictions show this behavior is expected to be different when $\gamma_\lambda \neq \gamma_H$.\cite{Kogan2002}  Since the coherence length $\xi$ has the same anisotropy as $H_{c2}$, the vortex core has a different shape determined by the anisotropy of $\xi$ as compared to the current distribution far from the core determined by the anisotropy of the penetration depth $\lambda$.  This mismatch costs energy, which - for a sufficiently large difference between $\gamma$'s - causes the energy of vortices parallel to \emph{ab} to exceed their value for vortices along \emph{c} for any uniaxial anisotropy, since both cores and current distributions are circular for this orientation.   Thus, for MgB$_2$, close to $T_c$ where $\gamma_H=\gamma_\lambda$, the theory predicts that the torque will tend to orient crystals with their $ab$ plane parallel to the direction of the applied magnetic field $H$, similar to the observation for the high $T_c$ superconductors, whereas at low temperatures where $\gamma_H \ne\gamma_\lambda$ a single crystal will orient with its $c-$axis along the magnetic field.
\begin{figure}[!]
\includegraphics[width=2.5 in]{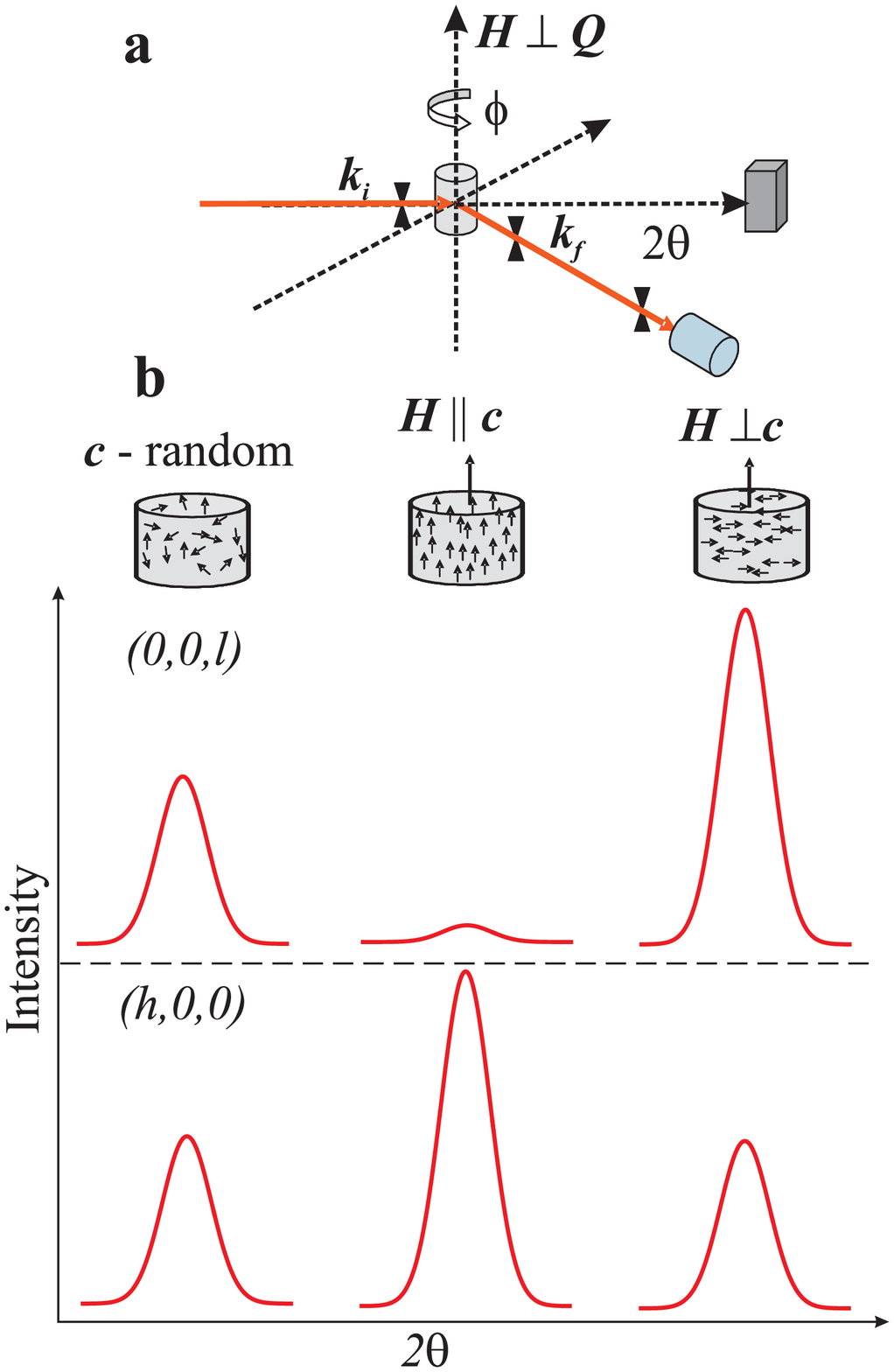}
\caption{\label{layout}
(Color online) a) Sketch of the diffractometer-magnet configuration.  The scattering vector is in the  plane normal to the direction of the applied magnetic field.  b) Illustration of the effect of single-crystal grains reorientations under magnetic field on the (\emph{0,0,l}) and (\emph{h,k,0}) reflections as a function of the scattering $2\theta$ angle.  The two sets of reflections at zero field (random orientation) are represented by same shape and intensity peaks for simplicity.  Preferred alignment of the $c-$axis  along the magnetic field, reduces the intensity of the (\emph{0,0,l}) increasing that of the (\emph{h,0,0}).  If the grains orient with the $ab$ plane parallel to the field $\bf{H}\perp\bf{c}$, the (\emph{0,0,l}) will become more intense and there will be only a slight effect on the intensity of the (\emph{h,0,0}).}
\end{figure}

 We have undertaken synchrotron X-ray diffraction studies to determine the orientation of free-grain powdered samples under applied magnetic field.  The preferential orientation can be detected in the X-ray scattering experiment on fine-grain MgB$_2$ powders provided a sufficient number of grains are single crystals and the difference in magnetic energy between the two principal orientations is large enough to overcome the average inter-grain energy barrier for reorientation.
 %, $E_f$.  Phenomenologically, we argue that the number of oriented crystals in a magnetic field ($N(T,H)$ is given by
 % \begin{equation}
 % N(T,H) = N_0\exp\left(-\frac{E_f}{\Delta\bf{M}\cdot\textit\bf{H}}\right)
 % \label{Pheno}
 % \end{equation}
% where, $N_0$ is the number of single crystals in the X-ray-illuminated volume.  (The inclusion of  Eq.\ (\ref{Pheno}) here is for later use in Fig.\ \ref{Field_dep} - I cannot justify it.)
In YBCO similar experiments found that the crystallites tend to orient themselves with the $c-$axis perpendicular to the applied field.\cite{Tranquada1988}
\section{Experimental details}
 The experimental setup used in this study and expected diffraction patterns for different scenarios are shown schematically in Fig.\ \ref{layout}.  The magnetic field is normal to the scattering plane spanned by the incident and scattered beams, $\bf{H}\perp\bf{Q}$; $\bf {Q \equiv k_i-k_f}$ is the scattering vector.  Figure\ \ref{layout} shows the consequences of applying a magnetic field on (\emph{0,0,l}) and (\emph{h,0,0}) reflections as a function of the scattering angle $2\theta$ represented with the same peak-shape and intensity at zero field (all grains are randomly oriented).  For simplicity, we assume the Bragg reflections are normalized to their structure factor, multiplicity, and Lorentz factor.  If the grains orient with their $c-$axis parallel to the applied magnetic field, we expect a near-complete suppression of the intensity of (\emph{0,0,l}) reflections and a significant increase in the (\emph{h,0,0}).  In the extreme case of perfect orientation ($\bf{H}\parallel\bf{c}$) and perfect instrumental resolution function ($\delta$-function-like) the (\emph{0,0,l}) reflections vanish.  In the case $\bf{H}\perp\bf{c}$, due to the axial symmetry, we expect a strong increase in intensity of the (\emph{0,0,l}) reflections and almost no change in the (\emph{h,0,0}), as depicted in Fig.\ \ref{layout}.

\begin{figure}[!]
\includegraphics[width=2.8 in]{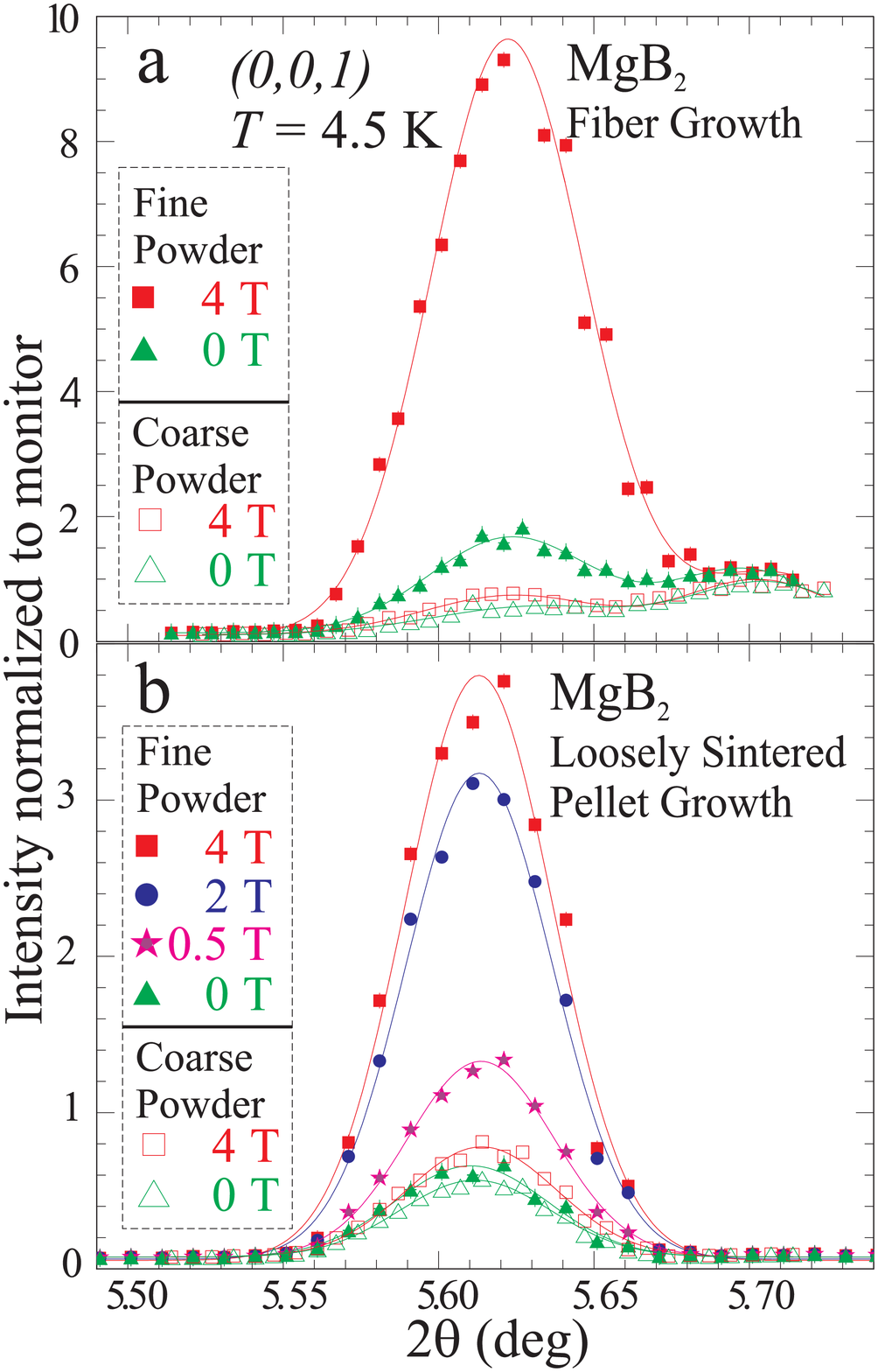}
\caption{\label{texture1}
(Color online) Intensity versus scattering angle $2\theta$ at $T = 4.5$ K of the  (\emph{0,0,1}) for types of MgB$_2$ samples a) fiber grown b) loosely sintered pellet grown  at various magnetic fields as indicated.  The intensity of each point is integrated over 60 degrees rotation of the sample angle $\phi$.  Whereas a significant increase in intensity is observed for finely ground powder (filled symbols), there is almost no change in intensity for samples consisting of coarse grains (open symbols). After lowering the magnetic field the powder was randomized spontaneously by mechanical vibrations of the compressor-cooled magnet with a typical time constant of a few minutes with small remnant alignment.\cite{comment1}}
\end{figure}

Two types of MgB$_2$ samples distinguished by the growth method and morphology, loosely sintered pellet\cite{Bud'ko2001}, and thin-wire segment (or fiber)\cite{Canfield2001}, were used in this study.   A portion of each sample was coarsely crushed to grains of 50 - 100 $\mu$m in size and another portion was thoroughly ground to 5 - 10 $\mu$m.  The four polycrystalline samples were loaded into thin walled (250 $\mu$m) cylindrical aluminum cans, and indium-sealed with copper caps in a helium environment to enhance thermal contact.  The X-ray scattering studies were conducted on the X-ray Operations and Research (XOR) beamline 4IDD, at the Advanced Photon Source, Argonne National Laboratory.   A Si(111) double crystal monochromator was used to select the 36 keV ($\lambda = 0.34440$ {\AA}) incident x-ray beam from an undulator 5th harmonic. A cryogen-free superconducting magnet with maximum field of 4 Tesla mounted on a Huber diffractometer was used to produce a magnetic field normal to the scattering plane as schematically shown in Fig.\ \ref{layout}.  The magnet is equipped with three optical beryllium windows providing access to a wide range of scattering angles.  A magnetically shielded scintillator point detector was used for measuring photon intensity.    The intensity of each point in $2\theta$ scans was integrated over 60 degrees of sample rotation $\phi$ to improve powder averaging, and normalized to a monitor count of the incident beam-intensity.  Samples were shaken thoroughly prior to being loaded in the cryo-magnet.  The superconducting magnet is cooled by a closed-cycle helium refrigerator that produces $\sim$ 1 Hz of continuous mechanical vibration of the sample stage.

\begin{figure}[!]
\includegraphics[width=2.8 in]{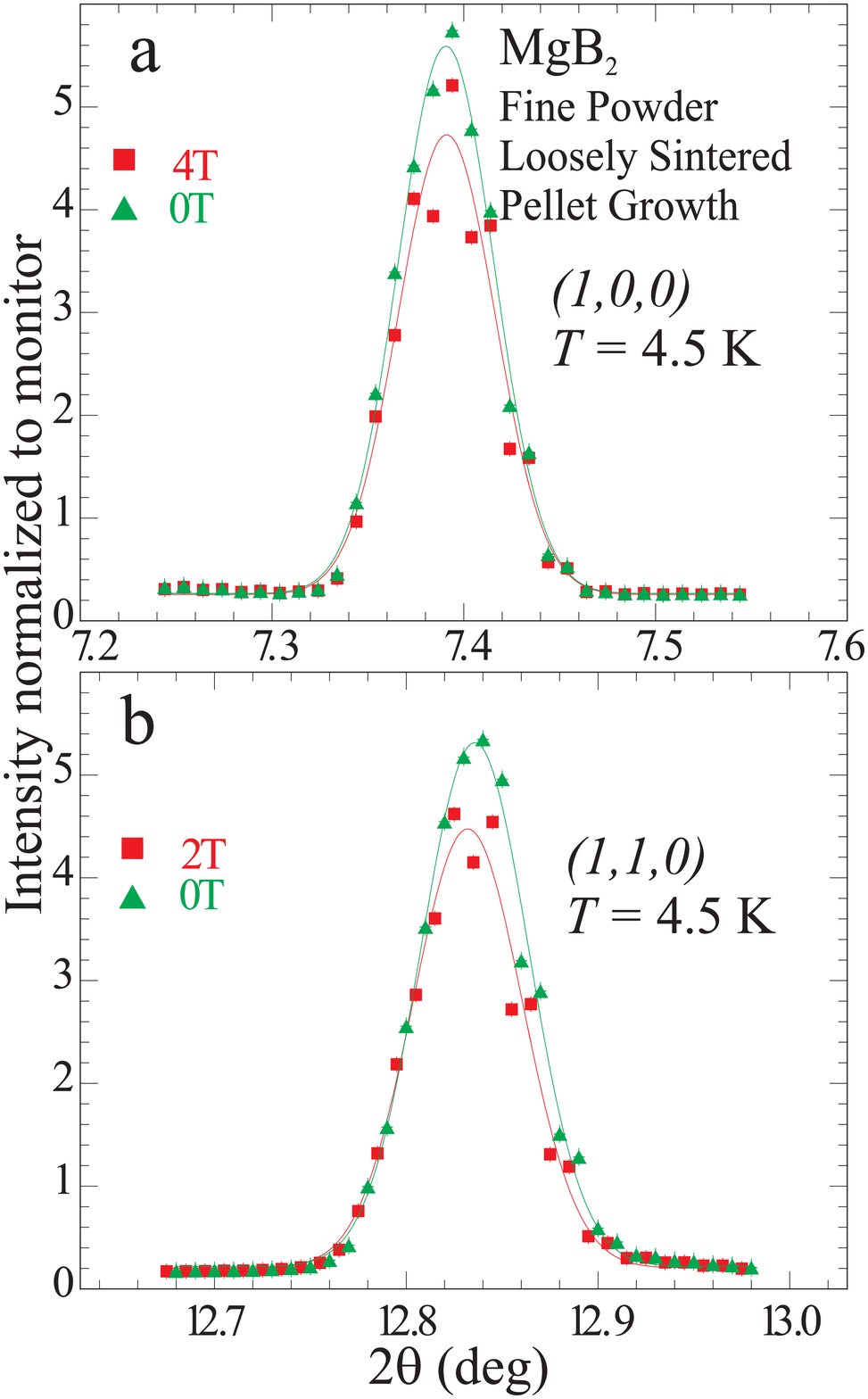}
\caption{\label{inplane}
(Color online) Intensity versus scattering angle $2\theta$ at $T = 4.5$ K  of (\emph{1,0,0}) and  (\emph{1,1,0}) in-plane Bragg reflection in magnetic field.}
\end{figure}
\section{Results and Discussion}
Figure\ \ref{texture1} shows intensity versus $2\theta$ scans of the (\emph{0,0,1}) Bragg reflection at $T = 4.5$ K demonstrating the effect of applied magnetic field on the coarse and fine powders of both the sintered pellet and the fiber grown.\cite{comment0}  Similar behavior with increase in intensity under applied magnetic field was also observed for the  (\emph{0,0,2}) Bragg reflection.   The coarse powders exhibit a 10 -20{\%} increase  between zero and the highest magnetic field (open symbols), whereas the increase in intensities for the fine powders is 7 - 8 fold stronger at 4 Tesla compared to zero field (solid symbols).  The field independent peak at $2\theta \approx 5.7^o$ is due to unidentified impurity phase present in the fiber grown sample.\cite{Canfield2001}   The difference in behavior of finely and course ground powders is in fact expected, as it is known that as grown MgB$_2$ even moderately ground consists of bound crystalline clusters and only thorough grinding yields isolated single crystals (5 - 10 $\mu$m in size).  By contrast to the (\emph{0,0,l}) reflections, the intensities of (\emph{h,k,0}) reflections yielded significantly smaller changes in intensity.   A very small decrease in the intensities of the inplane (\emph{1,0,0}) and (\emph{1,1,0}) reflections were detected, as shown in Fig.\ \ref{inplane}.  These observations clearly prove that the crystallites tend to orient with their $c-$axis normal to the magnetic field in the superconducting state.  After lowering the magnetic field to zero, the peaks intensities nearly recover their values in the randomized state within minutes after the field is removed.\cite{comment1}

\begin{figure}[ht]
\includegraphics[width=2.8 in]{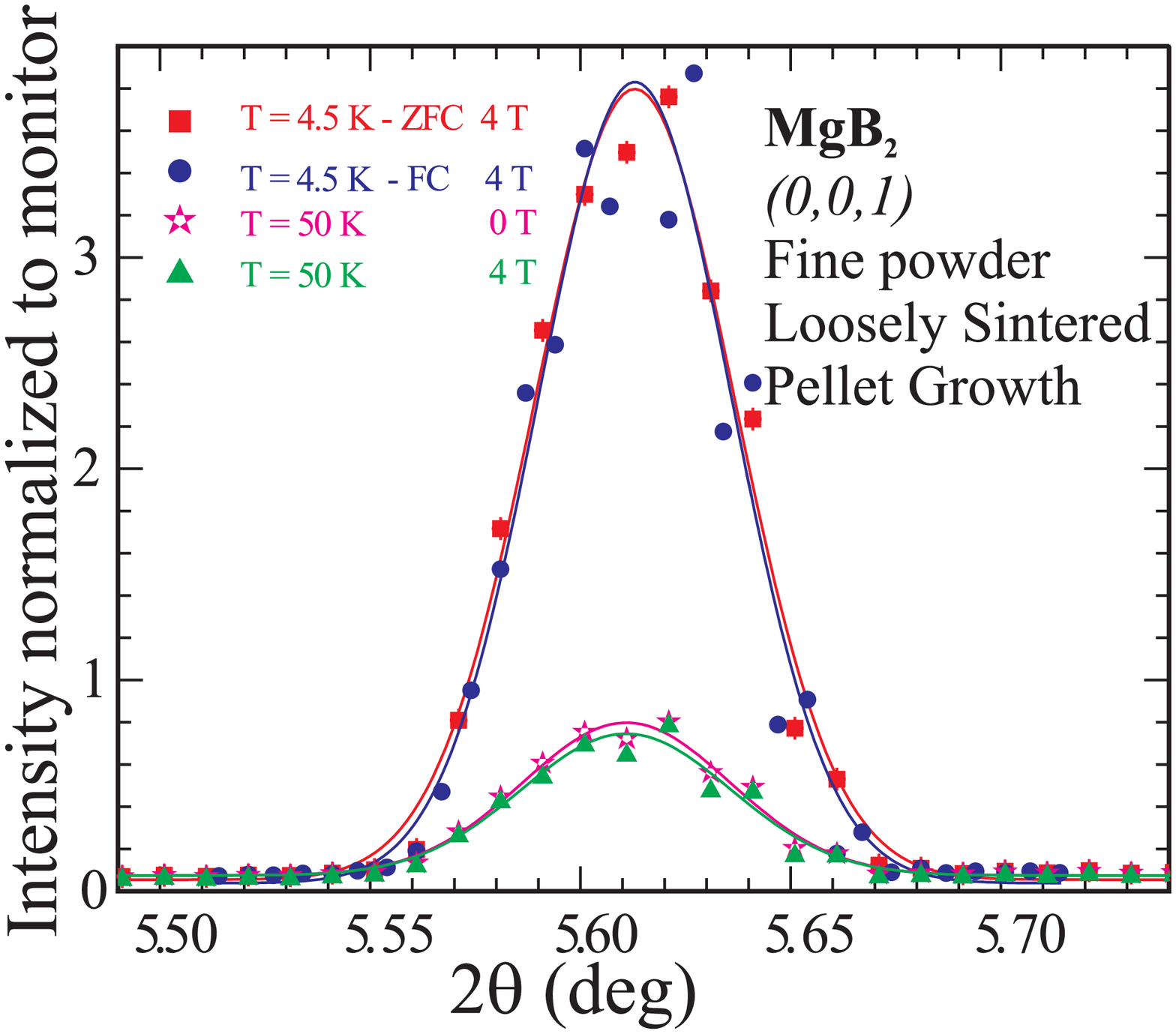}
\caption{\label{fc_001}  (Color online) Intensity versus scattering angle $2\theta$ at $T = 50$ K  at 0 and 4 Tesla of the (\emph{0,0,1}) of a finely ground sample.  The sample was then cooled in field to reproduce the results obtained for the zero-field cooled sample shown in Fig.\ \ref{texture1}(b).}
\end{figure}
The difference in the reorientation of field cooled (FC) or zero field cooling (ZFC) samples was monitored by the intensity of the (\emph{0,0,1}) reflection.   Figure\ \ref{fc_001} shows scans of the (\emph{0,0,1}) reflection for finely ground sample that was cooled under 4 Tesla from $T = 50 $ K to base temperature (4.5 K, circles) in comparison with a similar scan of a ZFC sample (square symbols).  There is no observable difference of FC or ZFC protocols on grain orientation.  Figure\ \ref{fc_001} also includes two scans of the (\emph{0,0,1}) reflection at $T = 50 $ K, at zero and 4 Tesla with practically no change in either intensity or shape. Hence, the orientation occurs only in the superconducting state.

\begin{figure}[!]
\includegraphics[width=2.8 in]{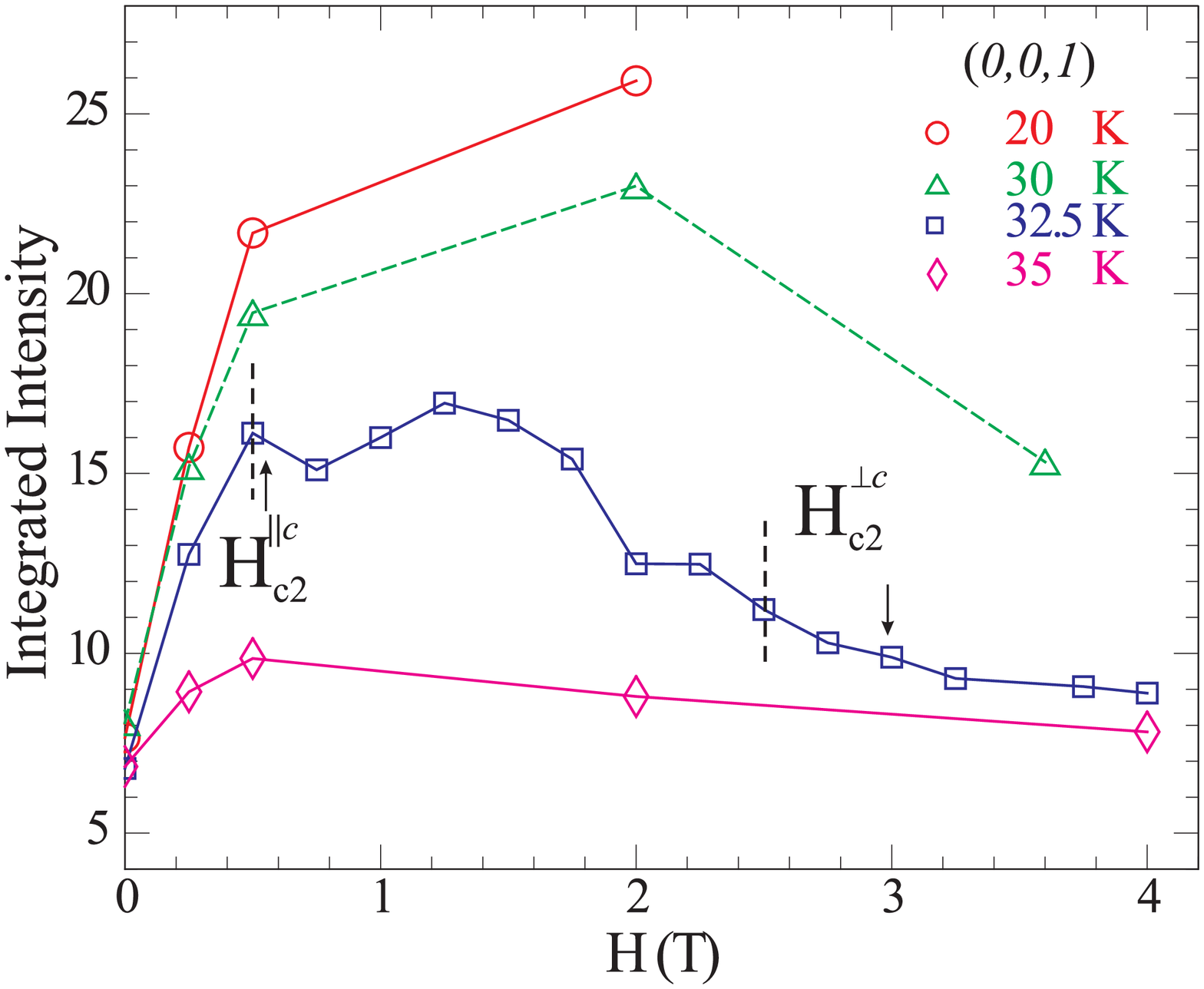}
\caption{\label{Field_dep}   (Color online) Integrated intensity of (\emph{0,0,1}) versus magnetic field at various temperatures, for the finely ground loosely sintered pellet grown sample.   We suggest to identify $H^{\parallel c}_{c2} \approx 0.5$ T as the point of deviation from the trend of the low field data, i.e. with development of the first minimum and to assign $H^{\perp c}_{c2} \approx 3$ T with the point of deviation from the asymptotic behavior at high fields.  Magnetization measurements of $H^{\perp c}_{c2}$ and $H^{\parallel c}_{c2}$ at 32.5 K on similar powders from Ref.\ \onlinecite{Bud'ko2001b} are shown with dashed lines. }
\end{figure}

The degree of crystalline orientation is both temperature and field dependent.  Figure\ \ref{Field_dep} shows the integrated intensity of the (\emph{0,0,1}) peak at various temperatures versus magnetic field.  At all temperatures the intensity of the (\emph{0,0,1}) initially increases with field providing evidence that the ground state orientation does not vary with temperature.  The more detailed field dependent measurement, conducted at $T = 32.5$ K (open squares), shows features that can be correlated with $H_{c2}^{\parallel c}$ T and $H_{c2}^{\perp c}$, indicated by arrows.  We suggest to identify $H^{\parallel c}_{c2} \approx 0.55$ as the point of deviation from the trend of the low field data, and with the development of the first minimum, and to identify $H^{\perp c}_{c2} \approx 3$ T with the point of deviation from the asymptotic behavior at high fields (the curves shown in Fig.\ \ref{Field_dep} for all temperatures other than 32.5 K are not sufficiently detailed to extract either  $H^{\parallel c}_{c2}$ or $H^{\perp c}_{c2}$).   It should be noted that above $H_{c2}^{\parallel c}$ the particles with the c-axis parallel to the field are in the normal state, and only by virtue of mechanical vibrations may tilt to become superconducting and  orient with the {\it c-}axis perpendicular to the field.  This explains the decrease in orientation at $H_{c2}^{\parallel c}$.  The dashed lines in Fig.\ \ref{Field_dep} are the results obtained from magnetization measurements.\cite{Bud'ko2001b}  Although the X-ray diffraction technique described here appears promising as a method to determine upper critical fields, more systematic studies are obviously required before it can be utilized as a reliable tool for this purpose.
In conclusion, our X-ray diffraction data show that the preferred orientation of an MgB$_2$ crystallite at low temperatures is to align with $c\perp H$ whereas theory predicts\cite{Kogan2002} the orientation should be $c||H$ orientation.   Furthermore, the orientation $c\perp H$ persists at all temperatures showing no evidence of orientation change.   One can trace the disagreement to the assumption that the anisotropy parameter $\gamma_{\lambda}$ and its temperature dependence used to interpret the high-field properties of the mixed state in Ref. \onlinecite{Kogan2002} is the same as the one  calculated\cite{Kogan2002b} and measured\cite{Feltcher2005} for small fields.  The theoretical calculation\cite{Kogan2002b} was done for a fixed ratio of gaps on two major sheets of the Fermi surface of MgB$_2$ ($\pi-$ and $\sigma-$bands), whereas by now it is common that the small $\pi-$gap is suppressed even in moderate fields of a few kOe.\cite{Eskildsen2003}.  Thus, in the fields used in this experiment, the material behaves as having only one gap at the $\sigma$-band and therefore a single anisotropy parameter for both $H_{c2}$ and $\lambda$ (i.e., $\gamma_\lambda$ is nearly the same as $\gamma_H$).  In other words, calculations of Ref.\,\onlinecite{Kogan2002} do not apply.  To construct a model applicable for all fields, one would need information on the angular dependence of the $\pi$-band suppression field. In principle, this can be acquired by detailed measurements of the field and angular dependencies of the magnetization and torque.  The field and temperature dependence of the alignment show features that can be associated with the $H_{c2}$ transitions yielding yet another way to measure the superconducting anisotropy in powder samples.

%\section{Results and Analysis}

%\subsection{\label{sec:level2}Second-level heading: Formatting}

% \subsubsection{\label{sec:level3}Third-level heading: References and Footnotes}
{\acknowledgements We wish to thank M. Tillman and D. Wilke for help in sample preparation.
The work at Ames Laboratory is supported by Basic Energy Sciences through the Ames Laboratory under Contract No.
W-7405-Eng-82.  Use of the Advanced Photon Source is supported by the U.S.
Department of Energy, Basic Energy Services, Office of Science, under Contract
No. W-31-109-Eng-38. MRE is supported by the Alfred P. Sloan Foundation.}


\begin{references}
\bibitem{comment0}  The intensities of the fine powders, fiber and the loosely sintered pellet, at zero field are different (Fig.\ \ref{texture1}) possibly due to differences in average grain size and scattering-volume filling ratios, however, the degree of alignment in the highest magnetic field (4 T) for both samples is similar, we therefore report on the loosely sintered pellet fine powder henceforth.
\end{references}
\end{document}